\documentclass{aastex631}

\usepackage{amsmath}	

\newcommand{\eq}[1]{\begin{equation}  #1 \end{equation}}
\newcommand{\br}[1]{\left( #1 \right)}
\newcommand{\vek}[1]{\mbox{\boldmath $#1$}}

\begin{document}
\title{An Extreme Radio Fluctuation of Pulsar B1929$+$10}
\author{Zhengli Wang}
\affiliation{Guangxi Key Laboratory for Relativistic Astrophysics, School of Physical Science and Technology, \\
Guangxi University, Nanning 530004, China}

\author{Shunshun Cao}
\affiliation{Department of Astronomy, School of Physics, Peking University, Beijing 100871, China}

\author{Jiguang Lu}
\affiliation{National Astronomical Observatories, Chinese Academy of Sciences, Beijing 100012, China}
\affiliation{Guizhou Radio Astronomical Observatory, Guiyang 550025, China}

\author{Yulan Liu}
\affiliation{National Astronomical Observatories, Chinese Academy of Sciences, Beijing 100012, China}
\affiliation{Guizhou Radio Astronomical Observatory, Guiyang 550025, China}

\author{Xun Shi}
\affiliation{South-Western Institute for Astronomy Research (SWIFAR), Yunnan University, Kunming 650500, China}

\author{Jinchen Jiang}
\affiliation{National Astronomical Observatories, Chinese Academy of Sciences, Beijing 100012, China}

\author{Enwei Liang}
\affiliation{Guangxi Key Laboratory for Relativistic Astrophysics, School of Physical Science and Technology, \\
Guangxi University, Nanning 530004, China}

\author{Weiyang Wang}
\affiliation{School of Astronomy and Space Science, University of Chinese Academy of Sciences, Beijing 100049, China}

\author{Heng Xu}
\affiliation{National Astronomical Observatories, Chinese Academy of Sciences, Beijing 100012, China}

\author{Renxin Xu}
\affiliation{State Key Laboratory of Nuclear Physics and Technology, Peking University, Beijing 100871, China}
\affiliation{Department of Astronomy, School of Physics, Peking University, Beijing 100871, China}
\affiliation{Kavli Institute for Astronomy and Astrophysics, Peking University, Beijing 100871, China}

\correspondingauthor{Jiguang Lu, Enwei Liang, Renxin Xu}
\email{lujig@nao.cas.cn, lew@gxu.edu.cn, r.x.xu@pku.edu.cn}
\begin{abstract}
We report the detection of an extreme flux decrease accompanied by clear dispersion measure (DM) and rotation measure (RM) variations for pulsar B1929+10  during the 110-minute radio observation with the Five-hundred-meter Aperture Spherical radio Telescope (FAST). 
%
The radio flux decreases by 2 to 3 orders of magnitude within a rapid time scale of about 20 minutes.
 %
 Meanwhile,  the variations of DM and RM are approximately 0.05 pc cm$^{-3}$ and 0.7 rad m$^{-2}$, respectively.
 %
 %
 Frequency-dependent analysis of DM indicates an extremely weak chromatic DM feature, which does not notably affect the radiative behavior detected.
 %
 Moreover, the pulsar timing analysis shows an additional time delay from 100\,$\mu$s to 400\,$\mu$s in the event.
 %
%
%
%
These results are speculated to be due to the eclipse and bend for the radio emission of pulsar B1929+10 by a highly dense outflow from the pulsar. 
This not only impacts the intrinsic radio emission feature but also affects the pulsar timing behavior. 
Nevertheless, a plasma lens effect lasting around 20 minutes could also be responsible for the event.
\end{abstract}
\section{Introduction} \label{sec0}
The interstellar medium can modulate the radio emission of the pulsars in various ways
such as the electron density fluctuations (inducing scintillation)
 and the scattering that can influence the detection of the radio pulsars \citep[e.g.,][]{1990ApJ...364..123F,2010arXiv1010.3785C}. These effects also influence the understanding of the intrinsic radio emission of the pulsars. Besides, some regions with higher electron density can cause plasma lensing phenomena \citep[e.g.,][]{1987Natur.326..675F,2016Sci...351..354B}. The origin of these events is usually interpreted as the extreme scattering events (ESEs). Theories of the plasma lensing effects based on the two different density profiles such as Gaussian and power-law cases to bend the rays of the targeted sources \citep[e.g.,][]{2017ApJ...842...35C,2020ApJ...892...89L,2022MNRAS.516.2218E}.
\par Previous observations of the plasma lensing events due to the ESEs, indicate that these plasma lensing events impact the amplitude of the flux density of the targeted radio source usually had about 50$\%$ decrease or a few events reached 10$^{-1}$ decrease. 
They usually last for long durations, ranging from several hours to days, and even several weeks to months \citep[e.g.,][]{1987Natur.326..675F,1993Natur.366..320C,2002PASA...19...10C,2016Sci...351..354B}. A short time scale of the plasma lensing event was observed by \cite{2018Natur.557..522M} in an eclipsing binary system. Due to this plasma lensing event, the observed radio flux is enhanced by factors ranging from 70 to 80 at specific frequencies.
\par In this work, we report an extreme radio fluctuation of pulsar B1929$+$10 that includes a decrease down to 10$^{-3}$ in the pulsar's flux density which lasted for about 20 minutes during our 110 minutes observation. The short duration and the low flux are unusual compared to previously reported plasma lensing events(due to the ESEs or in an eclipsing binary system). The variations in pulsar B1929$+$10 could help us to constrain the local region around this pulsar and the environment along the line of sight (LOS), but it is challenging to interpret their origins. These variations are important for studying the ISM environment and understanding the intrinsic radio emission of the targeted radio sources. Meanwhile, this event yields an additional delay in pulsar timing measurement, which influences the pulse arrival times with sub-microsecond accuracy required for the detection of nanohertz-frequency gravitational wave (GW) \citep[e.g.,][]{1990ApJ...361..300F,2006Sci...314...97K,2013ApJ...762...94D,2021PhRvX..11d1050K,2023RAA....23g5024X,2023A&A...678A..50E,2023ApJ...951L...8A}.
\par Section \ref{sec1} describes the 110 minutes observation for pulsar B1929$+$10 carried out by the FAST and data processing. In Section \ref{sec2}, we present the related results of this observation, while in Section \ref{sec3} we discuss the outflows from the pulsar that result in the variations in pulsar B1929$+$10 and present other possible origins. Section \ref{sec4} summarizes our main conclusions.
\section{Observation and DATA processing} \label{sec1}
The normal 226\,ms pulsar B1929$+$10 with a distance of $\sim$ 0.331\,kpc \citep[e.g.,][]{2002ApJ...571..906B,2004MNRAS.353.1311H} and with both a main pulse and an interpulse \citep[e.g.,][]{1990ApJ...361L..57P,1997JApA...18...91R}, was observed with FAST on MJD 59904 (November 21, 2022) using 19-beam receiver system \citep[][]{2019SCPMA..6259502J,2020RAA....20...64J}. The observation lasts 110 minutes and is carried out in tracking mode by FAST. The raw data were recorded in the 8-bit-sampled search mode PSRFITS format \citep[][]{2004PASA...21..302H} with a sampling time of 49.152\,$\mu$s. The frequency resolution is about 0.122\,MHz, and the recorded frequency channel number is 4096. To obtain precise polarization emission features of the averaged pulse and single pulse, accurate polarization calibration signals of 30\,s were injected after each observation length for 30 minutes required in our observation. The data processing was carried out by the \textsc{DSPSR} software package \citep[][]{2011PASA...28....1V}, and the raw data were folded with a time resolution of 0.226\,ms. About 28000 individual pulses were obtained from the entire 110-minute observation. The radio frequency interference (RFI) is mitigated by using the dynamics spectrum in the frequency-time to eliminate the impact of RFI on the pulse emission of pulsar B1929$+$10.
\section{Results} \label{sec2}
\begin{figure*}
    \begin{minipage}[t]{.55\textwidth}
        \centering{\includegraphics[width = 1\textwidth]{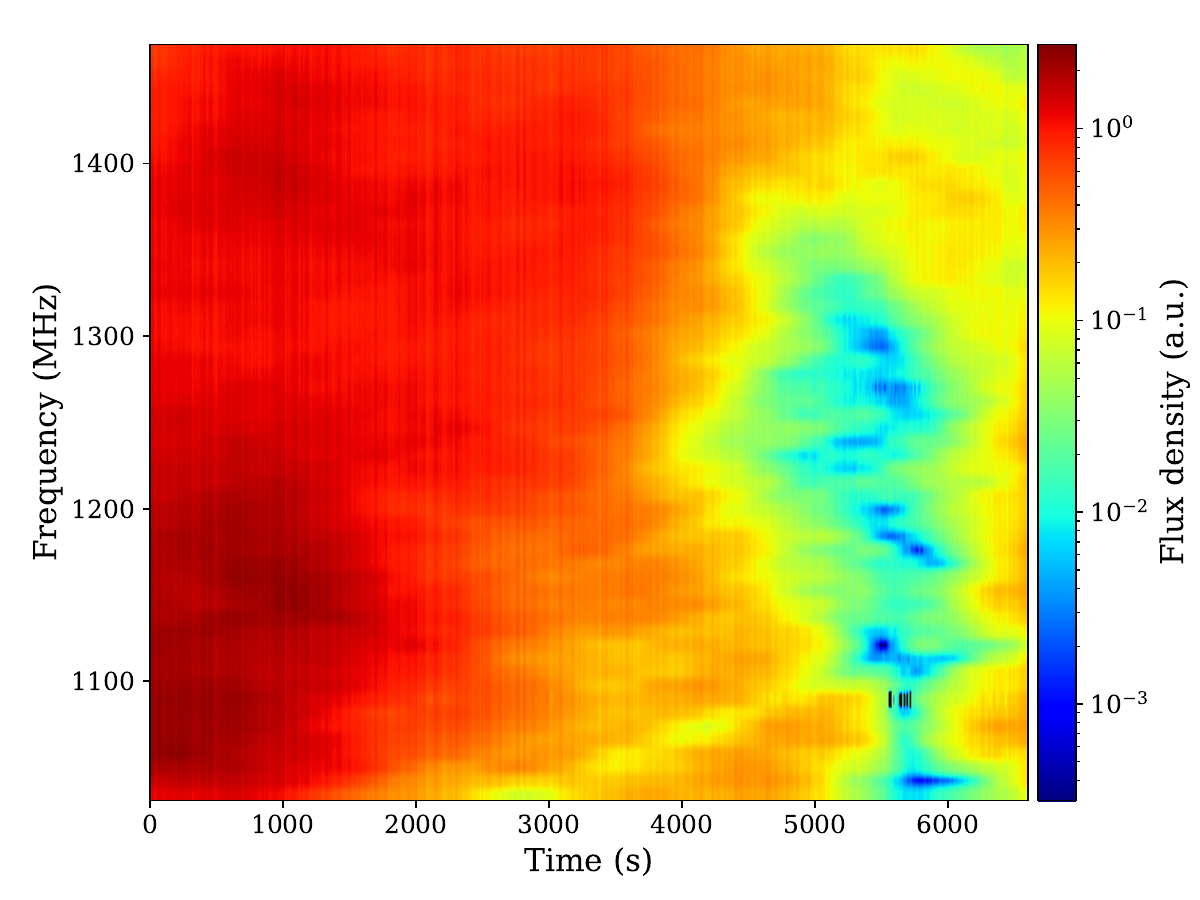}}
        \centerline{(a)}
    \end{minipage}
    \begin{minipage}[t]{0.45\textwidth}
        \centering{\includegraphics[width = 1\textwidth]{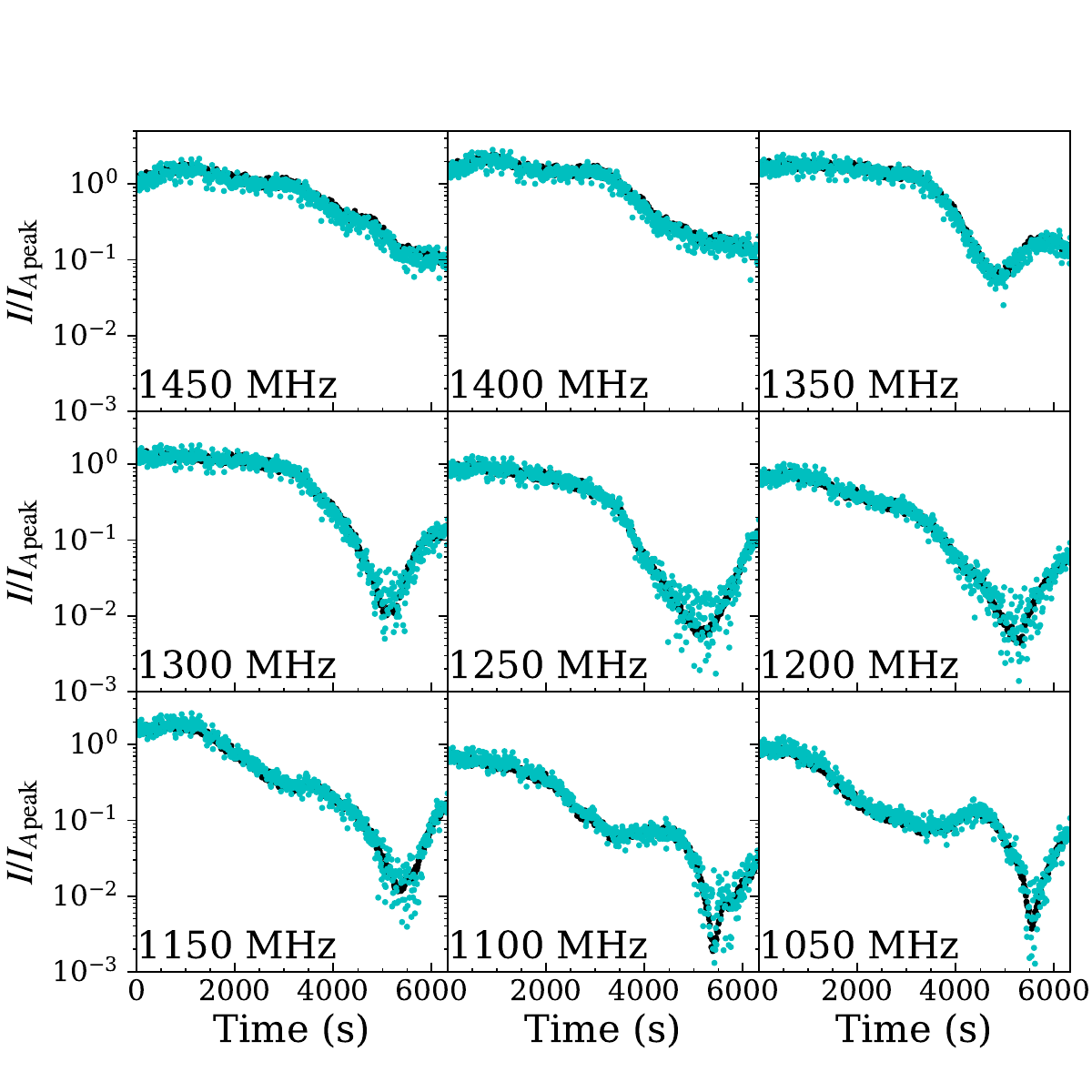}}
        \centerline{(b)}
    \end{minipage}
    \caption{(a) Dynamic spectrum averaged to around 8\,MHz frequency resolution for the main pulse of pulsar B1929$+$10. We choose an integration time of approximately 10\,s (45 rotation periods) for each spectrum to minimize the impact of system noise. The flux density has been scaled with the peak radio emission of the averaged pulse. (b) To better reveal variations over the frequencies and the arrival time. We plot the observed flux densities (black dots) for the main pulse as a function of the observation time for nine narrow bands with a bandwidth of 50\,MHz. Their corresponding centered frequencies are included in the lower left corner of each subpanel, while the cyan dots denote the flux density of the interpulse. The $I_A{_\mathrm{peak}}$ represents the peak radio emission of the average pulse for the main pulse, corresponding to the peak radio emission of the interpulse of the averaged pulse. The integration time of each dot is the same as that of each spectrum.}
    \label{flux}
\end{figure*}
\begin{figure*}
    \centering{\includegraphics[width = 1.\textwidth]{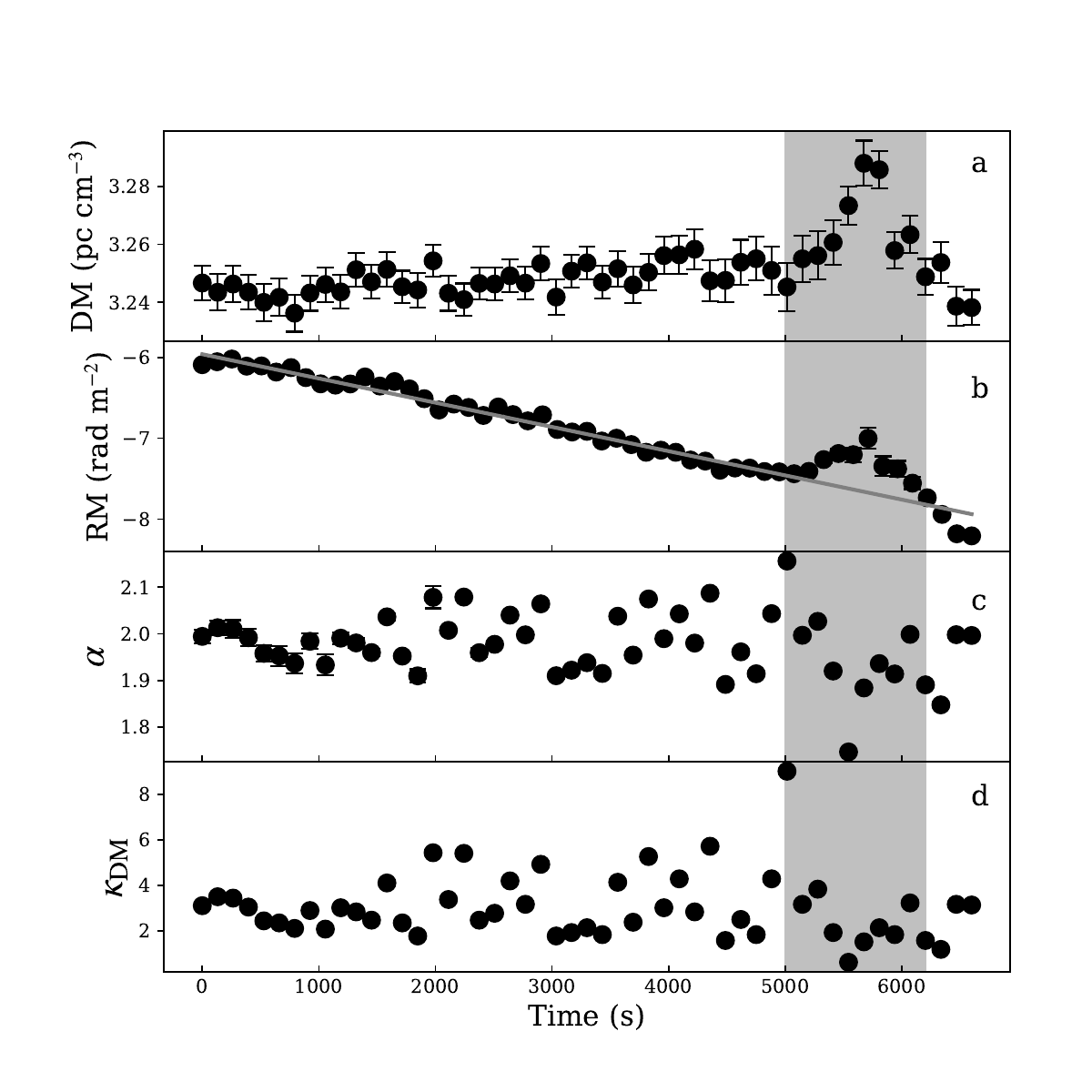}}
    \caption{Variations over the entire 110 minutes observation. We plot the measured DM (a), the measured RM (b), the detection of $\alpha$ (c), and the measured $\kappa_{\mathrm{DM}}$ (d), as a function of the observation time. Each dot corresponds to the integration of 2 minutes. To better exhibit the event, we use the vertical gray shadow to mark the region of the event.
    \label{dmrm}}
\end{figure*}
\begin{figure*}[ht!]
\centering
\includegraphics[width = 1.\textwidth]{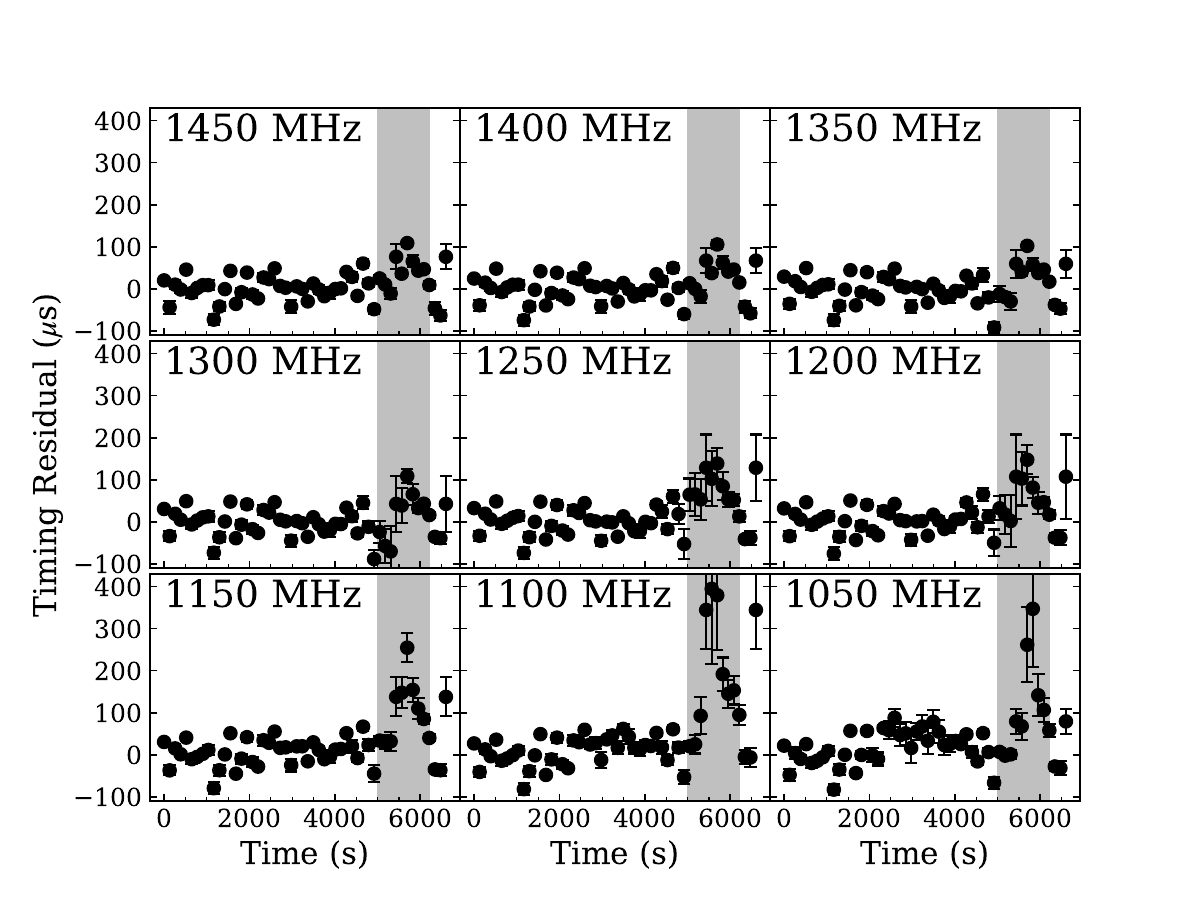}
\caption{Narrowband pulsar timing of nine narrow frequencies with a bandwidth of 50\,MHz as a function of the observation time for pulsar B1929$+$10, their corresponding frequencies are included in the upper left corner of each panel, as labeled. In each panel, the black dot represents the arrival time of a pulse profile in a short duration of 2 minutes.}
\label{toa_nb}
\end{figure*}
\begin{figure*}
    \centering{\includegraphics[width = 1.\textwidth]{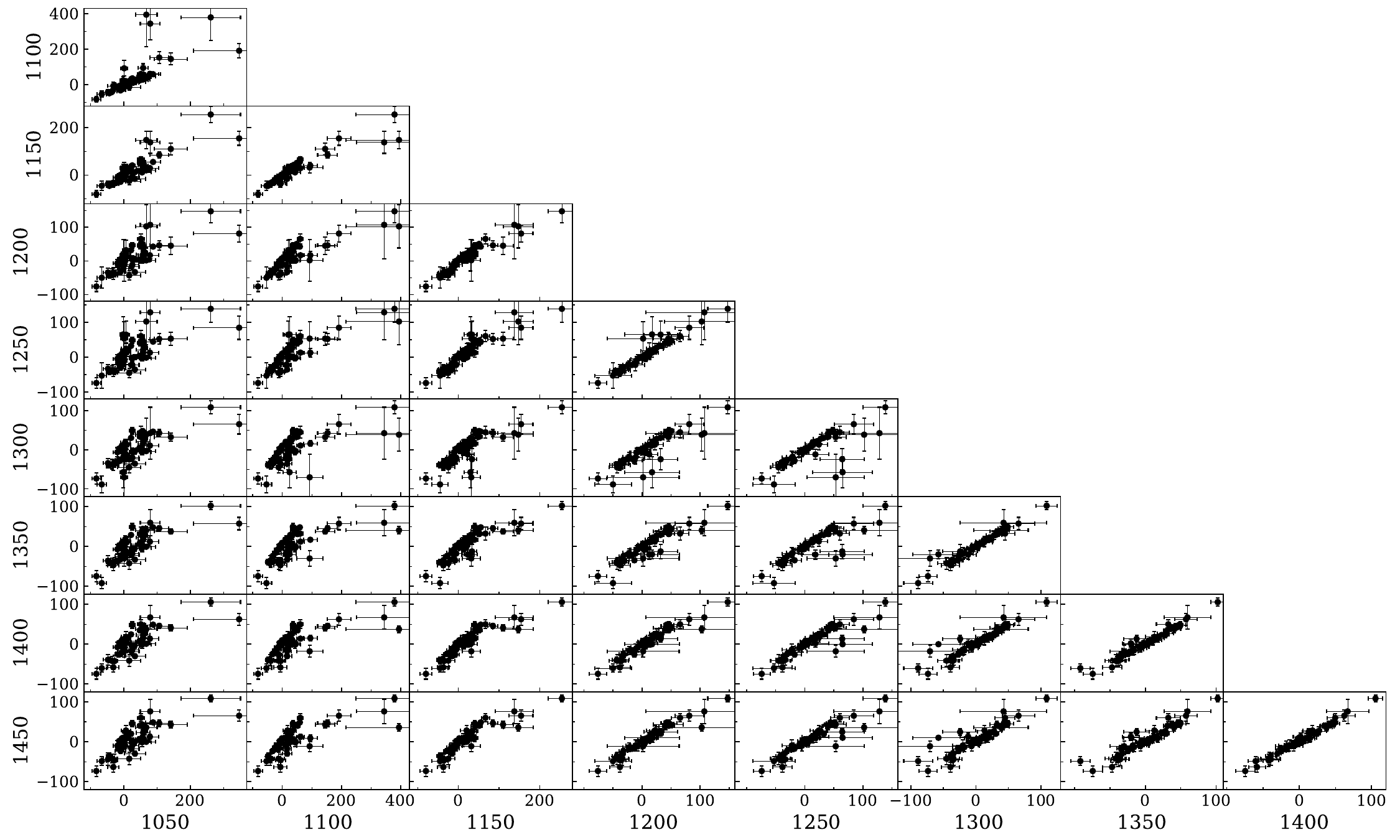}}
    \caption{The scatter plots of the timing residuals between pairs of frequencies. The frequencies are labeled and in units of MHz. The dot corresponds to the arrival time of a pulse profile in a short duration of 2 minutes and is in units of microseconds.}
    \label{pars_of_freq}
\end{figure*}
%
%
\begin{figure*}[ht!]
\centering
\includegraphics[width = 1.\textwidth]{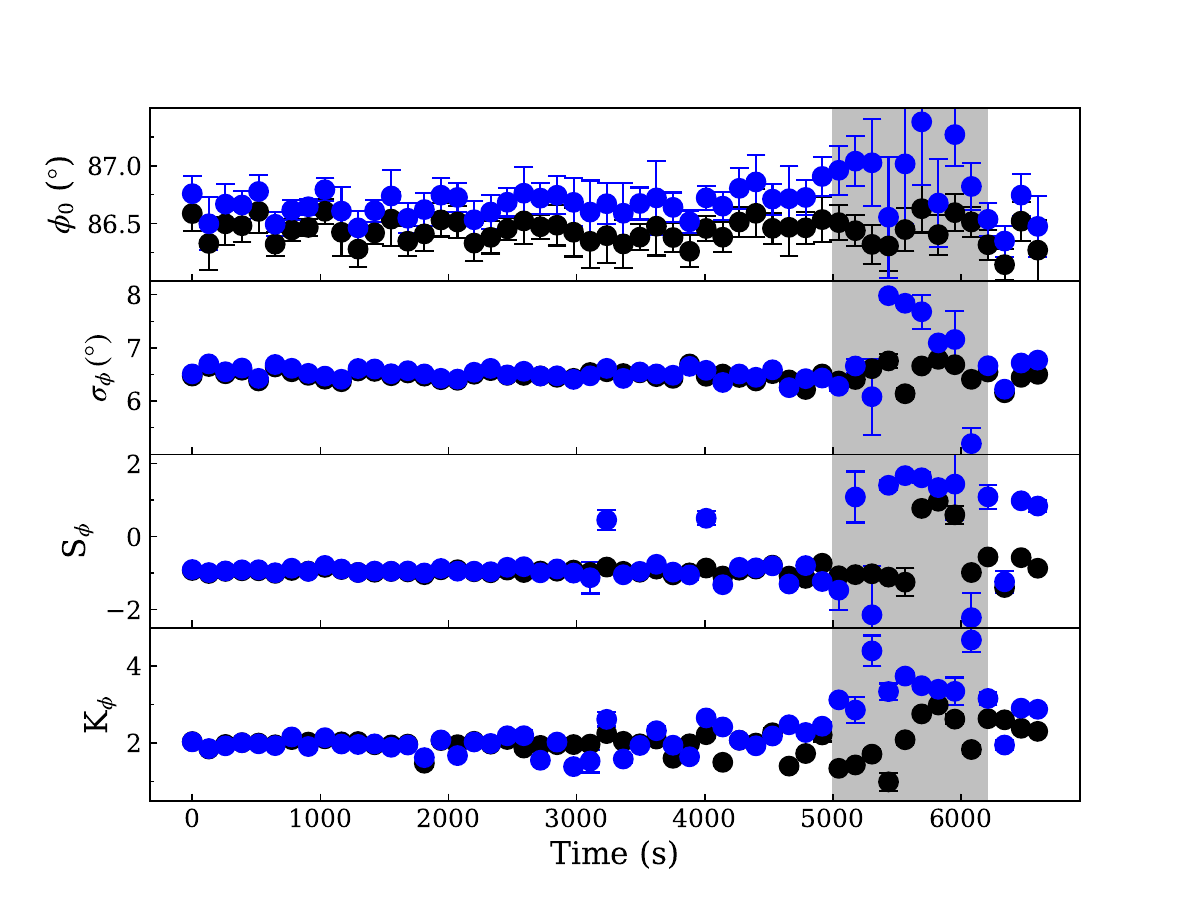}
\caption{The parameters are used to detect the behaviors of the profile defined in Equations (\ref{eq2}), (\ref{eq3}), (\ref{eq4}), and (\ref{eq5}) as a function of the observation time. The black dots represent the results of the observing frequency 1250\,MHz with a bandwidth of 450\,MHz, while the blue dots correspond to that of the frequencies that are lower than 1200\,MHz. The dot corresponds to the result in a 2-minute integration. The vertical gray shadow corresponds to the region of the event.
\label{psb}}
\end{figure*}
%
\subsection{Variations in the observed flux density}
Pulsar B1929$+$10 is a bright pulsar, so the variation in its observed flux density is a probe to unravel the environment along the LOS~\citep[e.g.,][]{1990ARA&A..28..561R,2022A&A...664A.116L,2024MNRAS.527.7568O}. Figure \ref{flux} depicts the fluctuations in the observed flux density of pulsar B1929$+$10, showing that the flux density exhibits a distinctive and significant fluctuation. These fluctuations depend on observing frequencies. Except for the frequencies higher than 1375\,MHz, the observed flux density at other frequencies shows a significant decrease of 3 orders of magnitude. After dropping to the minimum, the observed flux density then increases, forming a ``valley'' structure (the blue and navy-blue patches) during the range of 5000 to 6200\,s. The timescales of the decrease and increase are almost equal, implying a symmetrical ``valley'' structure. The decrease up to about 3 orders of magnitude and the maximum width of the ``valley'' (the time scale of the fluctuation) is detected at 1250\,MHz. The observed flux density at some frequencies decreases even more than 3 orders of magnitude (the navy-blue patches). There is an increase in flux density around 3200s at frequencies lower than 1125MHz, forming a shallow ``valley" structure (the yellow patches) from 2000 to 5000\,s. The observed radio flux is reduced by a factor of about 10. Moreover, the yellow patches experience a time delay as the frequency increases.
\par The start of the fluctuation in the observed flux density also shows a frequency difference. The variations start earliest from 4000\,s and are detected at 1250\,MHz. The dramatic decrease starting at 1100\,MHz is observed from 5500\,s. In addition, in Figure \ref{flux}(b) one can see that the arrival time of minimum flux density has a significant decrease when the frequency increases, showing a power-law function of $t = 1.36\times 10^{5}f^{-0.46}$. Here $t$ and $f$ are measured in units of second and MHz, respectively. Moreover, the observed flux densities at frequencies higher than 1375\,MHz decreased but did not increase over the left 90 minutes observation after the integration time of 1000\,s.  
However, this increase is not detected in other frequencies. Another decrease of the flux density was detected at 1350\,MHz after the observation of 5800\,s again.
\subsection{Variations in DM and the detection of the chromatic DM}
The dispersion measure in a cold, uniform, unmagnetized plasma yields the dispersive delay between different frequencies can be described as follows \citep[e.g.,][]{2012hpa..book.....L,2016ApJ...817...16C,2024A&A...683A.183M},
\begin{equation}
    \Delta t = \mathcal{D} \times \mathrm{DM} \times \left(\frac{1}{f_1^{2}} - \frac{1}{f_2^{2}}\right). \label{eq0}
\end{equation}
Here, $\mathcal{D} = \frac{e^2}{2 \pi m_e c}$ corresponds to the dispersion constant, $e$ and $m_e$ are the charge and mass of the electron, $c$ is the speed of light. As described by the relation $\mathrm{DM} = \int  n_e d l$, where $n_e$ represents the electron number density, and the integration $\int dl$ is along the line of sight. The radio waves of the pulsars can be believed to be a probe that unravels the environment along the LOS. Variations in DM indicate the electron number density is changed, modulating the radio wave of the pulsar along the LOS. We analyze the measurements of DM and RM to determine the variation in the environment over the 110-minute observation. In this work, we dedisperse the raw data in the frequency-time of 2-minute integration in a series of DM trials to obtain the pulse profiles. Then we calculate each profile's signal-to-noise ratio (SNR) based on each DM trial and determine the best DM as the one with the highest SNR in the profile.
Figure \ref{dmrm}a shows the variation in the measured DM in a short duration of 2 minutes, showing that a large increase in DM up to 0.05 pc cm$^{-3}$ is detected in the event. This detection implies a compact plasma environment appears along the LOS. Moreover, the changes in the measured DM results in an additional delay in the pulsar timing about 130\,$\mu$s at the frequency of 1250\,MHz.
\par To estimate the averaged parallel magnetic field along the LOS during the DM jumps, the rotation measure (RM) of pulsar B1929$+$10 is determined over the entire 110-minute observation. The RM was measured using the \textsc{LAPUDA} software package \footnote{\url{https://github.com/lujig/lapuda}}. We consider the diffusion in the Stokes ($Q, U$) space for each phase bin. Then, we fit the relation between the polarization position angle and the square of the wavelength, $\lambda^2$. In Figure \ref{dmrm}b one can see that there is a decreasing trend (the grey line $\mathrm{RM} = -3.0 \times 10^{-4} t - 5.96 $. Here RM and $t$ are measured in units of rad m$^{-2}$ and second, respectively.) in the measured RM. We utilize the ionFR software package \footnote{\url{https://github.com/csobey/ionFR}} to estimate the contribution of the earth's ionosphere, $\mathrm{RM}_{\mathrm{ion}}$, to the RM. We find the $\mathrm{RM}_{\mathrm{ion}}$ has no change during our 110 minutes observation. This suggests that the decrease with a linear trend in the measured RM throughout the 110 minutes has another origin. Besides, a significant increase up to 0.7\,rad m$^{-2}$ in RM is detected during the DM jumps. Considering that the relation $\mathrm{RM} \propto \int n_e B_z dl$, where $B_z$ corresponds to the field component parallel to the LOS. Both taking the variations in the DM and the RM into account, we can estimate the average parallel magnetic field along the LOS, $\left<B_z \right> = 12.3\, \mu \mathrm{G}\, (\Delta \mathrm{RM}/0.1 \, \mathrm{rad}\, \mathrm{m}^{-2}) / (\Delta \mathrm{DM}/0.01 \, \mathrm{pc} \,\mathrm{cm}^{-3}) = 17.22 \,\mu \mathrm{G}$, up to 17 times strengthen of the Galactic magnetic field (For the Galaxy, $\left<B_z \right> \sim 1 \,\mu \mathrm{G}$, see \cite{2012hpa..book.....L}), in the event. It suggests the presence of a magnetized environment with a highly compact plasma, which modulates the flux density of pulsar B1929$+$10 by a magnitude of 3 orders and causes changes in the measured DM and RM. Figure~\ref{dmrm}a also shows that the DM has a ramp with a positive slope variation before the event. 
%
\par The frequency-dependent (chromatic) DM is a stochastic component that is worth studying because the chromatic dispersion can influence radio wave propagation and yield an additional delay into the pulsar timing \citep[e.g.,][]{2016ApJ...817...16C,2020ApJ...892...89L}. To determine whether the event is affected by the chromatic dispersion, we quantify the total dispersive delay that includes the contributions of all possible DM components. Following \cite{2016ApJ...817...16C}, we consider the chromatic DM as a power-law function of the frequency $f$ (i.e., $\propto f^{\beta}$). The total dispersive delay resulting in the time difference between frequencies can be described as,
\begin{equation}
    \Delta t = \mathcal{D} \times \kappa_{\mathrm{DM}} \times \left(\frac{1}{f_1^{\alpha}} - \frac{1}{f_2^{\alpha}}\right), \label{eq1}
\end{equation}
where $\kappa_{\mathrm{DM}}$ denotes the total DM that includes all possible components. The $\kappa_{\mathrm{DM}}$ corresponds to the DM component described in Equation (\ref{eq0}) if the $\alpha$ is reduced to 2. While the $\alpha$ is a mixed index that describes the relation between the total dispersive effect and frequency.
\par We dedisperse the raw data in the frequency-time of 2-minute integration into a range of $\kappa_{\mathrm{DM}}$ and $\alpha$ trials to get the pulse profiles. We calculate the SNR of the pulse profile of each $\kappa_{\mathrm{DM}}$ and $\alpha$ trial based on the peak and the fluctuation in the ``pulse-off'' of the pulse profile. Similar to the pdmp software package \citep[][]{2011PASA...28....1V} that is used to search a specified range of barycentric period and DM of the pulsar. We consider the SNR as a function of the $\kappa_{\mathrm{DM}}$ and $\alpha$ and search the optimal values for $\kappa_{\mathrm{DM}}$ and $\alpha$ according to the highest SNR in the ($\kappa_{\mathrm{DM}}$ - $\alpha$) plane. Then we determine the best $\kappa_{\mathrm{DM}}$ and $\alpha$ and use them to dedisperse the raw data of all Stokes parameters ($I, Q, U, V$) in a 2-minute integration in the frequency-time domain. Figure \ref{dmrm}c and d shows the measurement of the chromatic DM in this observation, implying that there is no significant effect of the chromatic dispersion on the radio wave propagation of pulsar B1929$+$10. It is worth noticing that there is a slight deviation from 2 in the value of $\alpha$ detected after the observation time of 2000\,s, particularly in the range of 5000 to 6200\,s. This deviation is due to system noise and cannot be considered as evidence for the effect of chromatic dispersion.
\subsection{Measurement of the narrowband pulsar timing}
The time of arrival of the radio waves is not only used to detect nanohertz gravitational waves directly but also is believed to be a probe to reveal the environment along the LOS \citep[e.g.,][]{1986ApJ...310..737C,1995ApJ...443..209A,2004ApJ...605..759B,2014MNRAS.443.3752L}. The event shows an extreme fluctuation, decreasing in flux density by 3 orders of magnitude, encouraging us to analyze the pulse arrival time of pulsar B1929$+$10 in this observation. As Figure \ref{flux} shows, the characteristics of fluctuations in observed flux density display frequency differences, implying that radio waves of different frequencies are differently affected by the environment of the ISM. To accurately detect the properties of radio waves propagating multi-frequency paths, we analyze the narrowband timing of pulsar B1929$+$10.
\par Figure \ref{toa_nb} shows the timing residuals of pulsar B1929$+$10, it demonstrates that the radio waves have a significant time delay in the event. These timing residuals depend on the observing frequencies. The residual timing of the radio waves at the two lower frequencies (1100 and 1050\,MHz) have maximum values with an average of 300\,$\mu$s. The arrival time of the pulse display decreased feature with increased frequency, implying the pulsar timing behavior of pulsar B1929$+$10 can be interpreted as the DM increase. As the result is shown in Figure \ref{dmrm}a, change in the measured DM up to 0.05\,pc cm$^{-3}$ in the event. This increase can introduce a delay of about 130\,$\mu$s into the pulsar timing at 1250\,MHz. These results suggest that the radio waves of pulsar B1929$+$10 are affected by the environment, which is characterized by compact electron density along the LOS. 
In Figure \ref{pars_of_freq} one can see the highly corrected points between the adjacent frequency pairs (i.e., between 1100 and 1050\;MHz). This suggests that the timing measurements are caused by the DM increase. However, the correlation between pairs of frequencies below 1200\,MHz and those beyond 1200\,MHz does not show significant decreases as the frequency bands increase, implying that some other effects influence the timing measurements, such as the changes in pulse shape or the process of refraction. It is worth noticing that at large delay, the correlations diverge. This is because the radio emission flux at a frequency lower than 1200 MHz decreases the order of 3 in the event, resulting in large errors in timing measurements.
\subsection{Pulse shape}
To detect the pulsar timing behavior whether affected by the changes in the pulse shape, we analyze the profile in 2-minute integration throughout the 110-minute observation. We define the average pulse phase $\phi_0$ to detect the change of the profile,
\begin{equation}
    \tan \phi_0 = \frac{\left[\sin{\phi} \cdot I(\phi) \right]}{\left[\cos{\phi} \cdot I(\phi)\right]}\label{eq2},
\end{equation}
where $\phi$ and $I(\phi)$ correspond to the pulse phase with a range of $0$ to $2 \pi$ and flux density, respectively. To accurately detect the variation of the pulse shape intrinsic to pulsar B1929$+$10, we further define the standard deviation of pulse phase $\sigma_{\phi}$, the skewness of pulse phase $S_{\phi}$, and the kurtosis of pulse phase $K_{\phi}$ as follows,
\begin{equation}
    \sigma_{\phi}  = \sqrt{\frac{\sum_{\phi} \frac{\left[ \cos{(\phi - \phi_0)} + 1 \right]^{4}}{16} \cdot I(\phi) \cdot \left(\phi - \phi_0 \right)^2}{\sum_{\phi} I(\phi)}}, \label{eq3}
\end{equation}
\begin{equation}
    S_{\phi}  = \frac{1}{\sigma_{\phi}} \left(\frac{\sum_{\phi} \frac{\left[ \cos{(\phi - \phi_0)} + 1\right]^{4}}{16} \cdot I(\phi) \cdot \left(\phi - \phi_0 \right)^3}{\sum_{\phi} I(\phi)} \right)^{\frac{1}{3}}, \label{eq4}
\end{equation}
\begin{equation}
    K_{\phi}  = \frac{1}{\sigma_{\phi}} \left(\frac{\sum_{\phi} \frac{\left[ \cos{(\phi - \phi_0)} +1 \right]^{4}}{16} \cdot I(\phi) \cdot \left(\phi - \phi_0 \right)^4}{\sum_{\phi} I(\phi)} \right)^{\frac{1}{4}}. \label{eq5}
\end{equation}
Here $\phi_0$ denotes the average pulse phase, and $\frac{\left[ \cos{(\phi - \phi_0)} +1 \right]^{4}}{16}$ corresponds to the window function.
\par Figure \ref{psb} shows the measured values of the $\phi_0$, $\sigma_{\phi}$, $S_{\phi}$, and $K_{\phi}$ over the entire 110 minutes observation. For the results of the bandwidth of 450\,MHz, one can see that there are no variations in the $\phi_0$ and $\sigma_{\phi}$. 
A slight increases in the $S_{\phi}$ and $K_{\phi}$ are detected in the event. The slight changes of $S_{\phi}$ and $K_{\phi}$ may be caused by the effect of the system response since the flux density is decreased with an amplitude of 10$^{-3}$. These results suggest that there are no significant variations in pulse shape for the bandwidth of 450\,MHz over the entire 110-minute observation. However, for the results at a frequency lower than 1200\,MHz, significant increases in all of the parameters $\phi_0$, $\sigma_{\phi}$, $S_{\phi}$, and $K_{\phi}$ are detected in the event. The change of the $\phi_0$ is up to 0$^{\circ}.6$. This increase in $\phi_0$ corresponds to the delay of 370\,$\mu$s for pulsar B1929$+$10. This detection provides strong evidence that the pulse shape of the profile at a frequency lower than 1200\,MHz has changed. Consequently, the pulsar timing behavior shown in Figure~\ref{toa_nb} can be attributed to both the DM increase and the changes in the pulse shape effects.
\par The parameter $S_{\phi}$ is used to distinguish the effect of the chromatic dispersion and scattering on the radio wave propagation of pulsar B1929$+$10. If the radio wave from pulsar B1929$+$10 is affected by the scattering effect, the $S_{\phi}$ would have changed. The chromatic dispersion impacts on the radio wave of the pulsar yield an effect that can be regarded as the convolution between the radio waves and rectangular function. On the contrary, the scattering impact on the radio waves propagation can be equivalent to the convolution of the radio waves and a power-law function \citep[e.g.,][]{2006ApJ...637..346C}. These properties can be distinguished by the $S_{\phi}$ because the result due to the convolution of a rectangular function can not change the skewness feature of the samples. On the contrary, the effect owing to the scattering can alter the value of $S_{\phi}$. Figure \ref{psb} shows a $S_{\phi}$ change. This result indirectly implies that the scattering effect still impacts the radio wave propagation of pulsar B1929$+$10 in the event. This result is consistent with detecting the weak chromatic dispersion effect shown in Figure \ref{dmrm}c and d.
\section{Origin of the event} \label{sec3}
\subsection{The outflows near the pulsar}\label{sec31}
\par Based on the detection of the plasma lensing events due to the ESEs in previous observations \citep[e.g.,][]{1987Natur.326..675F,1993Natur.366..320C,2016Sci...351..354B}, these events are significantly different from the variations observed in pulsar B1929+10.
As shown in Figure~\ref{flux}, the observed flux density at a frequency lower than 1375\; MHz has a decrease of up to 3 orders, and at certain frequencies, the decrease exceeds 3 orders of magnitude, during the observation range of 5000 to 6200\,s. Meanwhile, the variations in flux density of pulsar B1929$+$10 are seen here as frequency dependence. For those frequencies whose flux densities show the ``valley'' structure, the arrival times of the bottom of the ``valley'' structure have increased with the decreased frequency (see Figure \ref{flux}(b)). These behaviors are hardly interpreted as the plasma lensing events caused by the ESEs.
\par
\begin{figure*}
    \begin{minipage}[t]{.5\textwidth}
        \centering{\includegraphics[width=1.\textwidth]{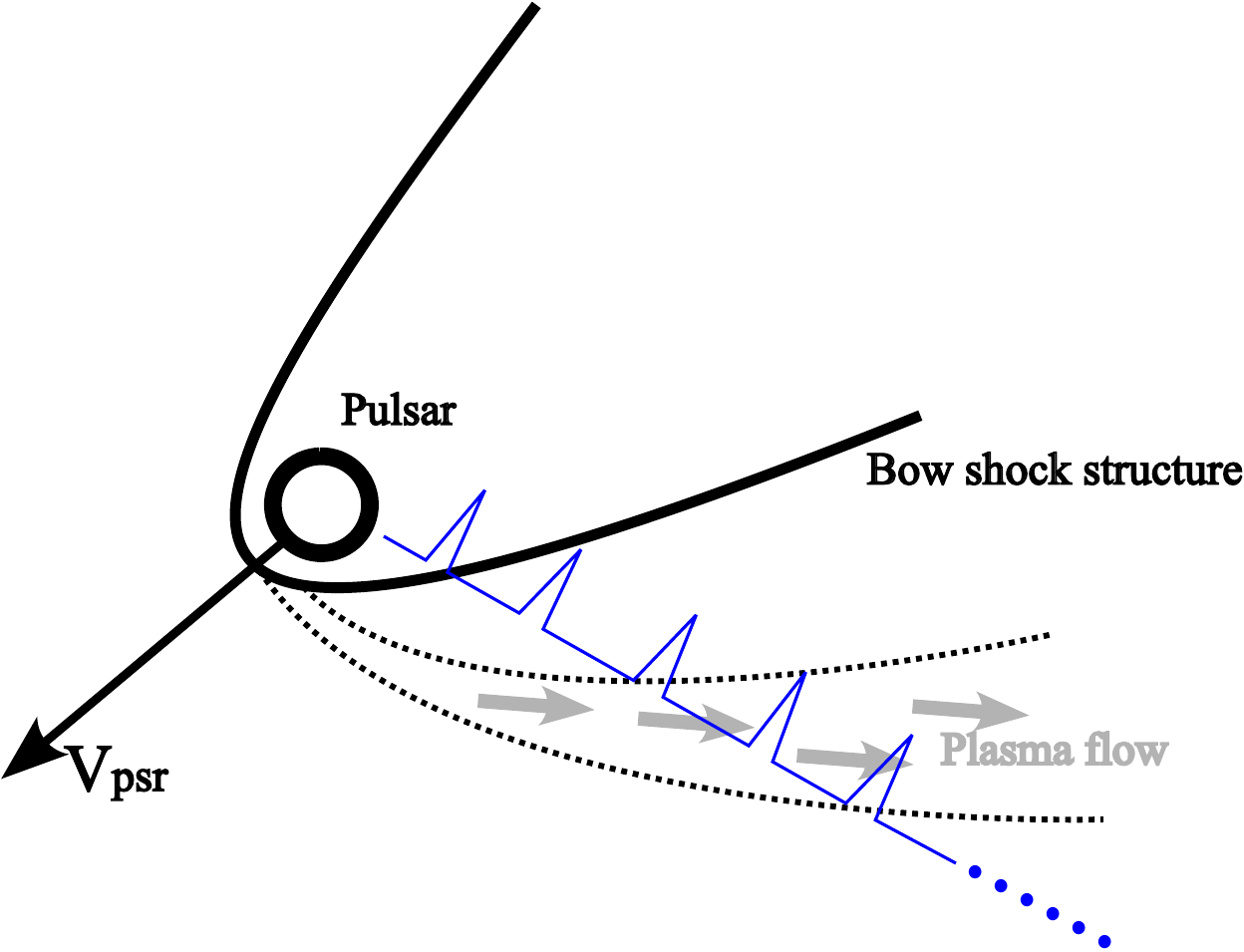}}
        \centerline{(a)}
    \end{minipage}
    \begin{minipage}[t]{.5\textwidth}
        \centering{\includegraphics[width=1.\textwidth]{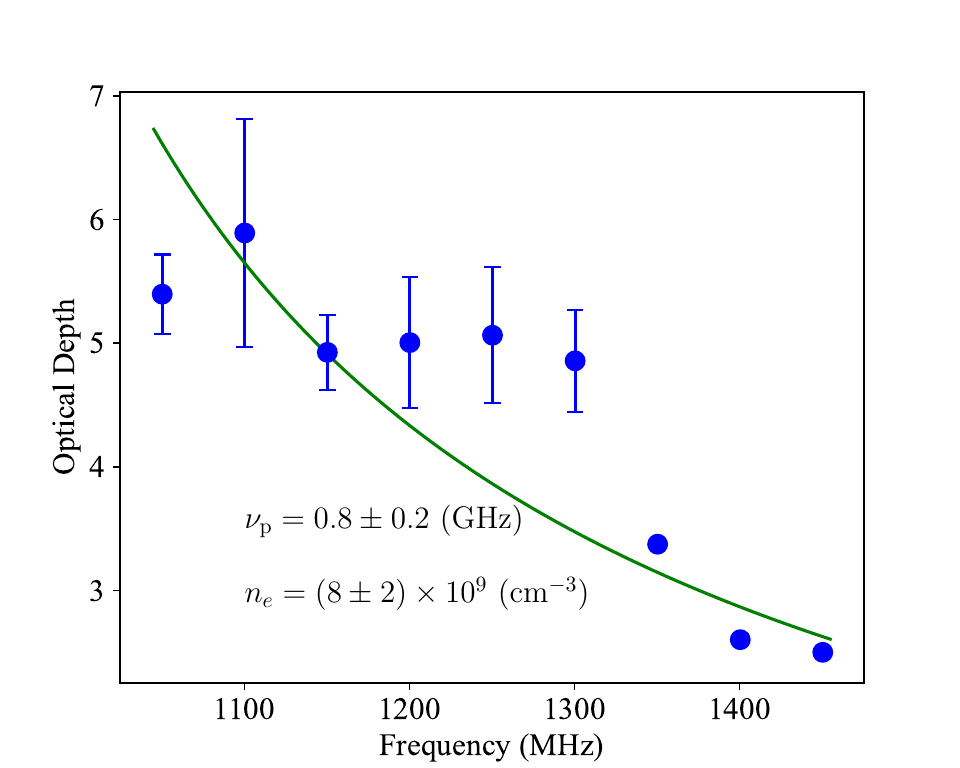}}
        \centerline{(b)}
    \end{minipage}
    \caption{(a) Sketch for the outflow from the pulsar. To better unravel the local environment near the pulsar, the bow shock structure and the proper motion are also included. (b) Fitting of optical depths under nine frequency bands using bremsstrahlung absorption. The optical depths are estimated from Figure~\ref{flux} through $\tau=\ln(I/I_{min})$ and are dotted as blue dots with error bars in the plot. The errors are estimated from the fluctuation in flux density in the ``pulse-off'' regions. The green curve of $\tau_{\mathrm{em},\nu}^{ff}$ is derived from the least square fitting. The derived $\nu_{\mathrm{p}}$ and $n_{e}$ are also marked on the plot.}
    \label{fig:brem_fit}
\end{figure*}

 The X-ray studies of a bow shock structure around pulsar B1929$+$10 imply that there exists an axial outflow that is opposite to its proper motion. This axial outflow appears to be longer than 4 arcmin~\citep[e.g.,][]{1993Natur.364..127W,2006ApJ...645.1421B,2020A&A...637L...7K,2024MNRAS.527.7568O}. Moreover, a deep X-ray spectral imaging was carried out by~\cite{2020A&A...637L...7K}, revealing two new lateral outflows around this pulsar. The outflows eclipse the radio wave of pulsar B1929$+$10, resulting in variations in this pulsar such as the flux density having an extreme decrease up to 10$^{-3}$ and the changes in the pulsar timing. In addition, the outflows can bend the rays, causing the plasma lensing process to increase the flux density, which is significant at 1050\,MHz during the range of 2000 to 4000\,s. The variation in $\Delta \mathrm{DM}$ of approximately 0.05\,pc cm$^{-3}$ detected in the event can help us constrain the region responsible for the plasma lensing. According to Equation 7 in \cite{2017ApJ...842...35C}, this increase yields the focal distance $d_f$ is less than 1\,pc. The value of $d_f$ is evidence of the region of the plasma lensing close to pulsar B1929$+$10. Furthermore, the region near the pulsar could be larger, resulting in a short timescale and yielding a larger focal distance. A schematic diagram of the eclipse and bending process is shown in Figure~\ref{fig:brem_fit}(a), the outflows eclipse and bend the rays of pulsar B1929$+$10 and then result in the radiative behaviors of pulsar B1929$+$10 within a short time scale (about 20 minutes) seen in this observation. To be sure, the physical origin of outflows is unclear, but it might not be surprising for active astrophysical objects, as in the case of solar energetic particle (SEP) events~\citep[][]{2023AdSpR..72.5161W}.
\par The variations in the flux density of pulsar B1929$+$10 can help us further estimate the outflows' parameters. As Figure~\ref{flux} shows, the flux density decreases by 3 orders of magnitude, and the maximum time scale of the modulation is detected at 1250 MHz. If consider that our observation frequency is near the plasma frequency $\nu_{\mathrm{p}}$ (i.e. $\nu_{\mathrm{p}} \approx 1250$\,MHz), then according to the relation $\nu_{\mathrm{p}} = \sqrt{ \frac{e^2 n_e}{\pi m_e}}$ \citep[][]{2012hpa..book.....L}, the electron number density is up to 2$\times 10^{10}$ cm$^{-3}$. This electron density and the increase in $\Delta \mathrm{DM}$ can estimate the outflows have a size of approximately 77\,km. Assuming that the outflows consist of electrons and protons, and the emission altitude of the normal radio pulsars is about 1000\,km, the mass of the outflows is up to about $250 \times 10^{3}$\,kg. The outflow just ejects from this pulsar since an extremely high electron number density can not last long in the ISM. However, the increase of $\Delta$RM is about 0.7 rad m$^{-2}$ in the event, which indicates a low magnetic field. It is challenging to understand the highly dense outflow that only has a low magnetic field. Besides, the low magnetic field speculated from $\Delta$RM$/\Delta$DM is typical for pulsar wind nebulae that are characteristic with low magnetization~\citep[e.g.,][]{2017SSRv..207..175R}. But by now we have no idea on whether the blocking particles are relativistic or not.
\par The electron number density of the outflows could also be estimated by assuming a major absorption/scattering mechanism that causes the optical depth $\tau$. The mechanism should be manifestly more effective at low frequency than at high frequency, as shown in Figure~\ref{flux}, so typical scattering processes should be ruled out. Besides, the small RM variation corresponds to the low magnetic field, so absorption mechanisms related to the magnetic field (cyclotron/synchrotron absorption) are less favored. We consider bremsstrahlung absorption (free-free absorption) as the major absorption mechanism, which is typical for astrophysical plasmas~\citep{1996ASSL..204.....Z}. The absorption coefficient of bremsstrahlung absorption at frequency $\nu$ for electromagnetic waves is~\citep{1996ASSL..204.....Z}:

\begin{equation}
    \alpha^{ff}_{\mathrm{em},\nu}=\dfrac{\nu_{{p}}^{2}\nu_{\mathrm{eff}}}{\nu^{2}c\sqrt{1-\nu_{\mathrm{p}}^{2}/\nu^{2}}}
\end{equation}
where $\nu_{\mathrm{p}}=\sqrt{4\pi n_{e}e^{2}/m_{e}}/2\pi$ is the plasma frequency, and $\nu_{\mathrm{eff}}$ is the collision frequency that depends on electron number density $n_{e}$, temperature $T$, etc. To give some constraints on the blocking flow's parameters, we try to fit the maximum optical depths of the nine frequency bands with an optical depth function $\tau^{ff}_{\mathrm{em},\nu} = \alpha^{ff}_{\mathrm{em},\nu}l$ assuming a uniform electron distribution. The least-square fitting result is shown in Figure~\ref{fig:brem_fit}(b). The plasma frequency derived is close to the observing frequency, and the electron density is about $10^{9}$ - $10^{10}$ cm$^{-3}$.

%
\subsection{Plasma lensing}
\par Usually, flux modulation on time scales of minutes to hours is attributed to interstellar scintillation, i.e., radio wave deflection by the ISM and interference of the scattered light. This particular observation is special because there is a large amplitude modulation with a broad frequency bandwidth on a relatively large time scale of $\sim$ 100 minutes at 1\,GHz, and the minimum radio flux is about 10$^{-3}$ of the average flux. This could be interpreted in terms of plasma lensing if the main image of the pulsar is broken into a small number of images of comparable brightnesses and small angular separations.

As an order-of-magnitude estimation, let us consider a simplified picture of the interference pattern created by two images separated by $a$ on the lens plane, with $a$ matching the lens size. 
Then, the width of the fringe pattern created on the observer's plane is
\eq{
    \lambda D_{\mathrm{l}} / a  \sim 0.3 \ \mathrm{m} \times 0.331 \ \mathrm{kpc} (1-s)/ a \sim 3\times 10^{22}\ \mathrm{cm^2} (1-s) / a
}
where $\lambda$ is the wavelength, $D_{\mathrm{p}}$ is the pulsar distance, the lens distance $D_{\mathrm{l}} = D_{\mathrm{p}} (1 - s)$, and $s$ is the fractional distance between the pulsar and the lens, we have taken $D_{\mathrm{p}} = 0.331 $ kpc~\citep[e.g.,][]{2002ApJ...571..906B}.

Taking the reference frame of a static pulsar-lens LOS, the relative velocity between the observer and this LOS is $\vek{v} = (1-s)/s \vek{V}_{\rm p} + \vek{V}_{\rm earth}$.
At the time of this observation, $V_{\rm earth, ra} = -18.3$ km/s, and $V_{\rm earth, dec} = 10.4$ km/s, whereas $V_{\rm p, ra} = -183.3$ km/s, and $V_{\rm p, dec} = 63.2$ km/s.
Thus, the amplitude of the relative velocity is $v = 165\sqrt{s^2-0.72s+1.23}/s$ km/s. 
Assuming the angle between the modulation direction of the fringe pattern and $\vek{v}$ is $\alpha$, then 
the width of the fringe pattern on a $\Delta t = 100$ minute scale is
\eq{
    v \Delta t \cos\alpha \sim \cos\alpha \sqrt{s^2-0.72s+1.23}/s  \times10^{11} \ \rm{cm} \,.
}

Combining these two expressions, we get
\eq{
    a \sim \frac{3 \times 10^{11}}{\cos\alpha} \frac{s(1-s)}{\sqrt{s^2-0.72s+1.23}} \ \mathrm{cm} \sim 3 \times 10^{11}\frac{s(1-s)}{\cos\alpha}\ \mathrm{cm} \,.
}
This corresponds to a sub-AU scale length unless the relative velocity is almost perfectly perpendicular to the fringe pattern.

If the DM increment is due to the plasma lens creating the fringe pattern, then the implied electron density enhancement in the lens 
\eq{
    \Delta n_{\rm e} = \Delta \mathrm{DM} /a/A \sim 0.05 \cos\alpha/A \ \mathrm{pc \;cm^{-3}} / \br{3s(1-s) \times 10^{11}\ \mathrm{cm}} \sim 5\times 10^5 \frac{\cos\alpha}{As(1-s)} \;\mathrm{cm}^{-3} 
}
is very high unless the aspect ratio $A$ of the lens (the ratio of its line-of-sight to transverse scales) is very large. Such small size and high density are not typical of the ISM and are possibly connected to the pulsar wind.

We have considered the additional conditions $\kappa > 1$ and $\kappa \tilde{\nu} > 1$ for plasma lensing to be responsible for the large-scale, large-amplitude modulation. Here the lensing convergence $\kappa$ represents a characteristic dimensionless lens amplitude. The reduced frequency $\tilde{\nu} = (a/r_{\mathrm{F}})^2$ with $r_{\mathrm{F}}$ being the Fresnel scale is a normalizing quantity proportional to the observing frequency such that $\kappa \tilde{\nu}$ is the phase shift induced by the lens \citep[e.g.,][]{2021MNRAS.506.6039S,jow23}. Whereas $\kappa$ being greater than unity is required for multiple images to exist, $\kappa \tilde{\nu} > 1$ is required so that radio wave effects do not smear out the flux modulation. However, both conditions can be satisfied at all $s$ values and thus give no constraint on the screen distances.

\subsection{An extra celestial object and other origins}\label{sec32}
\par An extra celestial object with a compact electron density atmosphere (brown dwarf or giant planet) transversely passes through the LOS to pulsar B1929$+$10 can also be responsible for this event. Its appearance causes variations in the flux density of this pulsar. Moreover, its atmosphere provides an environment with a magnetized and compact electron density to yield the DM and RM jumps when the radio wave of pulsar B1929$+$10 passes through it. Figure \ref{flux} shows the variations in flux density of this pulsar, which can help us to constrain the celestial object that has a diameter of about 2.4 $\times 10^5$\,km since the flux ``valley'' structure (the radio emission modulated by the celestial object) only lasts approximately 20 minutes. Here we consider the velocity dispersion estimated with the Virial theorem near the solar system to be approximately 200 km/s. 
This celestial object can be additionally verified through optical observations. Unfortunately, \cite{2002ApJ...580L.147M} reported the optical observation of pulsar B1929$+$10 and found this pulsar has optical emission behavior. In addition, we have tried to find a star near our LOS to pulsar B1929$+$10. According to VizieR Catalogue Collection\footnote{\url{https://vizier.cds.unistra.fr/viz-bin/VizieR}}, the minimum angular distance from observed stars to pulsar B1929$+$10 is 1.896 arcsec, corresponding to a Gaia source\footnote{id: 4314712501956058752}\citep{2022yCat.1355....0G}, but lack of distance measurements. Besides, the small RM variation is still a problem.
\par It is still worth noticing the possibility of the solar elongation of this pulsar at the time of our observation. It could be another possibility that is responsible for the radiative behaviors of pulsar B1929$+$10. Figure~(\ref{dmrm})a and b shows that both changes in the DM and RM were detected before the event. These phenomena also indicate that this pulsar’s motion through its local environment can account for some of the changes.
\section{conclusions}\label{sec4}
\par As Figure \ref{flux} shows the variations in radio emission flux of pulsar B1929$+$10 show frequency difference. Only frequencies below 1375 MHz show the radio emission behavior with a ``valley'' structure in the flux. The variations in the flux density at the frequencies below 1150\,MHz (the yellow patches with a positive slope in Figure \ref{flux}(a)), especially at 1050\,MHz, show that the observed flux density is reduced by a factor of 10, during the observation range of 2000 to 4000\,s, could be interpreted as the plasma lensing event. Moreover, the event shows that the flux density is modulated by the decrease of 3 orders of magnitude, forming a ``valley'' structure within a rapid time scale of about 20 minutes, this modulation is first detected in the radio emission behaviors of the pulsars.
\par Figure \ref{dmrm}a and b shows that both the DM and RM experience a jump during the event, indicating an environment that causes a decrease in flux density and an increase in electron density. The DM increase can cause an additional increase in RM. For pulsar B1929$+$10, the 0.05 pc cm$^{-3}$ increase yields the RM has 0.2 rad m$^{-2}$ increase. This RM increase due to the DM increase is 0.2 rad m$^{-2}$ is lower than the measured RM increase up to 0.7 rad m$^{-2}$. This deviation implies an additional magnetized environment, which modulates the magnetic field component parallel to the LOS of pulsar B1929$+$10. 
Moreover, the increase in RM in the event is positive going. That is the opposite of the negative ramp in RM and the positive ramp in DM before the event.
The radio emission behavior detected here can be useful for directly detecting chromatic DM. Following~\cite{2016ApJ...817...16C}, we investigate the effect of the chromatic dispersion and analyze the behavior of the radio wave of pulsar B1929$+$10. Figure \ref{dmrm}c and d show the detection of the chromatic DM. Variations in the index $\alpha$ and chromatic DM are detected after the observation time of 2000\,s, but these increases may be due to the narrowband frequency introducing correlation noise into the data and the system noise, and they are weak evidence for the detection of the chromatic DM. 
\par Pulse times of arrival of the pulsars are believed to be a probe to detect weak GWs with high precision timing since the GW can introduce systematic time delay into the pulsar timing measurement \citep[e.g.,][]{1989ApJ...345..434T,2006Sci...314...97K,2023RAA....23g5024X,2023A&A...678A..50E,2023ApJ...951L...8A}. We analyze the pulse's arrival time of pulsar B1929$+$10 to understand whether the radio waves are affected by the flux density event. Figure \ref{toa_nb} shows the pulsar timing results for the narrowband. It implies a significant time delay is detected in the event, consistent with a dispersive delay caused by the increase in DM of up to 0.05 pc cm$^{-3}$. These results provide strong evidence of a significant environmental change along the LOS of pulsar B1929$+$10. The behavior of the pulse shape shown in Figure~\ref{psb} indicates that the pulsar timing behavior of pulsar B1929$+$10 can also be attributed to the changes in pulse shape in the event.
\par There are three outflows around this pulsar were observed by~\citep[e.g.,][]{2020A&A...637L...7K}. As Figure \ref{fig:brem_fit}(a) shows, the outflows eclipse the radio waves of pulsar B1929$+$10 and bend its ray, yielding the plasma lensing effect that can be responsible for the above variations in pulsar B1929$+$10. The outflow may be just from this pulsar since it has high electron number density up to 10$^{10}$\,cm$^{-3}$ (The result is estimated from the variations in the flux density or the bremsstrahlung absorption). However, the outflow with a low magnetic field of approximately 17\,$\mu$G is still a challenge.
In addition, as discussed in Section~\ref{sec32}, an extra celestial object that transverse passes through the LOS to pulsar B1929$+$10 in the event is still possible. Its appearance could account for the detection of most of the variations. Also, the non-lensing physical process is possible, for example, this pulsar's motion through its local environment can contribute to some of the changes. Moreover, the effect of the solar system, such as that related to the variation of the pulsar's solar elongation, can not be completely ruled out.
\begin{acknowledgments}
{\em Acknowledgments.}
This work is supported by the Strategic Priority Research Program of the Chinese Academy of Sciences (No. XDA0350501, XDB0550300), the National SKA Program of China (No. 2020SKA0120100), the National Natural Science Foundation of China (12133003), the Guangxi Talent Program (“Highland of Innovation Talents”), and the Major Science and Technology Program of Xinjiang Uygur Autonomous Region, Grant No. 2022A03013-2 (Yulan Liu).
We are greatly indebted to Prof. James M. Cordes for reading carefully the manuscript and for his insights about radio waves traveling in the interstellar medium.
We would also like to thank Kejia Lee, Junzhi Wang, Haozhu Fu, and Xinyu Zhu for their useful discussions. This work made use of the data from the Five-hundred-meter Aperture Spherical radio Telescope (FAST), operated by the National Astronomical Observatories, Chinese Academy of Sciences.
\end{acknowledgments}

\bibliography{sample631}{}
\bibliographystyle{aasjournal}



\end{document}